\documentclass{article}

\usepackage{PRIMEarxiv}

\usepackage[utf8]{inputenc} 
\usepackage[T1]{fontenc}    
\usepackage{hyperref}       
\usepackage{url}            
\usepackage{booktabs}       
\usepackage{amsfonts}       
\usepackage{nicefrac}       
\usepackage{microtype}      
\usepackage{lipsum}
\usepackage{fancyhdr}       
\usepackage{graphicx}       
\graphicspath{{media/}}     

\usepackage{tikz}
\usepackage{subcaption}
\usepackage{amsmath,amsfonts}
\usepackage{algorithmic}
\usepackage[english,ruled,lined]{algorithm2e}
\usepackage{graphicx}
\usepackage{textcomp}
\usepackage{xcolor}
\usepackage{multirow}
\usepackage{booktabs}
\usepackage{pifont}

\usepackage{pgfplots}
\usepackage{pgfplotstable}
\pgfplotsset{compat=1.16}

\title{Network Performance Estimator with Applications to Route Selection for IoT Multimedia Applications
}

\author{
  Fabiano Bhering, Diego Passos, Célio Albuquerque \\
  Universidade Federal Fluminense (UFF) \\
  Niterói - RJ - Brazil\\
  \And
  Katia Obraczka \\
  Univesity of California, Santa Cruz (UCSC) \\
  Santa Cruz - CA - USA \\
}

\begin{document}
\maketitle

\begin{abstract}
Estimating the performance of multimedia traffic is important in numerous contexts, including routing and forwarding, QoS provisioning, and adaptive video streaming. This paper proposes a network performance estimator which aims at providing, in quasi real-time, network performance estimates for IoT multimedia traffic in IEEE 802.11 multihop wireless networks. To our knowledge, the proposed multimedia-aware performance estimator, or MAPE, is the first deterministic simulation-based estimator that provides real-time per-flow throughput, packet loss and delay estimates while considering inter-flow interference and multi-rate flows, typical of multimedia traffic.
Our experimental results indicate that MAPE is able to provide network performance estimates that can be used by IoT multimedia services, notably to inform real-time route selection in IoT video transmission, at a fraction of the execution time when compared to stochastic network simulators. When compared to existing deterministic simulators, MAPE yields higher accuracy at comparable execution times due to its ability to consider multi-rate flows. 
\end{abstract}


\section{Introduction}\label{sec_Introduction}
Efficient transmission of multimedia traffic in multihop wireless networks poses significant challenges mainly due to their more stringent Quality of Service (QoS) requirements (e.g., throughput and delay), especially in the case of real-time applications~\cite{Hasan2017-survey}. Additionally, multihop wireless communication is inherently more prone to losses and congestion; for instance, the performance of a single wireless link can vary due to factors such as link-layer transmission rate, its signal-to-noise-ratio (SNR), and complex propagation phenomena. Furthermore, transmission of multiple flows that are not limited by rate control mechanisms can also cause congestion, as well as inter-flow interference, medium access contention, and collisions~\cite{Ke2016-interflow}. And, in the specific case of multimedia traffic, even though compression techniques use a pre-defined average data rate as a target, the actual data rate of the compressed flow may vary considerably depending on scene complexity, flow resolution, and the different types of frames~\cite{seeling2011}.

Estimating network performance is an effective mechanism to address the challenges raised by multimedia traffic as a way to achieve QoS-aware admission control, resource provisioning and allocation in multihop wireless networks~\cite{reddy2006}. It allows estimating current available network capacity as well as deciding whether the network can fulfill each flow's requirements. In addition, accurate multimedia performance estimates are useful for routing and video coding decisions~\cite{bhering2022,Hasan2017-survey,wei2019evaluation}.

There is a wide variety of IoT (Internet of Things) multimedia applications that can benefit from a real-time network performance estimator to route selection~\cite{bhering2022}, such as surveillance systems for outdoor or indoor spaces in smart cities that require multiple video sources transmitting simultaneously to the monitoring center~\cite{Dao2018}. Note that the performance of these application scenarios can vary according to the selected route for each video flow, as this may cause inter-flow interference.

As will be discussed in more detail in  Section~\ref{sec_RelatedWorks}, different performance estimators have been proposed but do not fulfill the needs of IoT multimedia applications which require estimators to account for multi-rate flows as well as  inter-flow interference, while being able to provide their estimates in a timely and resource-efficient manner.

In this paper we propose the Multimedia-Aware Performance Estimator, or MAPE for short, which estimates network performance for multi-rate multimedia flows using their video coding rate as input. To the best of our knowledge, MAPE is the first estimator that is able to provide throughput, packet loss and delay estimates in real time considering rate-heterogeneous flows and accounting for inter-flow interference.  

Experiments using different IoT multimedia application scenarios demonstrate that MAPE is
able to provide real time network performance estimates, i.e., throughput, delay, and packet loss, with savings of over two orders of magnitude in execution time when compared to the \textit{ns-3}~\cite{riley2010ns3} network simulator. 
Furthermore, we show how MAPE can be used to improve video transmission quality by guiding route selection on a per-flow basis.

The remainder of this paper is organized as follows: Section~\ref{sec_RelatedWorks} reviews related work on network performance estimation in IEEE 802.11 networks. Section~\ref{sec_MAPE} describes MAPE's design and operation in detail. Our experimental methodology, and results from our evaluation of MAPE's accuracy 
are reported in Sections~\ref{sec_Experiments} and ~\ref{sec_Results}, respectively. Section~\ref{sec_Pratical} shows how MAPE can be used to guide route selection in order to improve video transmission quality. Finally, Section~\ref{sec_Conclusion} concludes the paper and presents directions for future work.

\section{Related Work}\label{sec_RelatedWorks}

IEEE 802.11 networks have offered several attractive rate-capable amendments that serve various multimedia application scenarios~\cite{Hasan2017-survey}. Providing performance estimates is critical to meet QoS guarantees in such networks. Existing approaches to network performance estimation in IEEE 802.11 networks can be classified in three main categories, namely: mathematical models, online estimators, and discrete-event simulators.

\noindent
\textbf{Mathematical models}
Estimators based on mathematical models typically make simplifying assumptions to make modeling tractable. For instance, most existing proposals target one-hop flows~\cite{wang2005throughput,garetto2008modeling,liew2010back}. Moreover, they make additional simplifications, such as perfect links and identical transmission rates for all nodes. In the context of per-flow performance estimation, Laufer and Kleinrock~\cite{laufer2015capacity} present a more complete model for analyzing the throughput of CSMA/CA networks. This model estimates the maximum throughput for each flow by modeling the network behavior as a system of non-linear equations and solving the resulting optimization problem. That approach can become prohibitively expensive for larger networks, as the size of the system of equations grows exponentially with the number of network nodes.

\noindent
\textbf{Online Estimators}
While mathematical models for performance estimation are useful to understand the limits of contention-based medium access protocols, approaches that can be operated online are required in practice, e.g., for real-time applications 
such as adaptive video streaming~\cite{bentaleb2019,jiang2014,sun2016cs2p,wei2018trust,biernacki2017improving,karn2019measuring,nabi2019,wei2019evaluation,wang2019} and routing protocols~\cite{campista2008routing}. In particular, performance estimation for adaptive video streaming is discussed in~\cite{wei2018trust,karn2019measuring}. These studies also consider buffer occupancy information for predicting performance to improve video streaming quality of experience (QoE). The work reported in ~\cite{wang2019} proposes a method to reduce the impact of inaccurate throughput prediction on QoE by controlling the buffer occupancy within a safe range. In turn, routing metrics provide indirect information that is expected to correlate well with throughput~\cite{campista2008routing}, but they usually fail to evaluate the interference between flows.

\noindent
\textbf{Discrete-Event Simulators}
Discrete event simulators can be stochastic or deterministic. 
Stochastic simulators use pseudo-random number generators to determine the outcomes of events that have some level of randomness (e.g., the choice of backoff intervals for medium access), while deterministic simulators replace pseudo-random generation  with deterministic values (e.g., a fixed average backoff interval).

Network performance estimation performed by stochastic simulators like \textit{ns-3}~\cite{riley2010ns3} and OMNET++~\cite{varga2001} is commonly used to either conduct an a-priori evaluation of a certain network and its protocols, guide network provisioning, deployment or operational tasks. 
Because of their random nature, they usually require a large enough number of runs for every experimental configuration in order to obtain statistically meaningful results, which adds to their inherent scalability limitations, long execution times, and high computational resource needs. 
On the other hand, deterministic estimators provide an adequate accuracy with identical results no matter how many times they are run. However, they must be designed to perform in real-time while the network operates to help dynamically adjust operational parameters.

One notable example of this latter class of performance estimators is AFTER~\cite{passos2018after}. It was proposed to tackle the problem of real-time throughput estimation for multihop IEEE 802.11 networks. AFTER simulates the behavior of the link- and network layers to quickly converge to steady state behavior that allows it to estimate the long term average throughput of each flow for a given set of application flows and corresponding routes. To this end, it maintains in memory a complete view of the network topology and performs a deterministic simulation of the network dynamics, generating simulated virtual packets (\textit{v\_packets}) for each flow at their respective virtual source nodes, triggering a number of other relevant simulation events, such as wireless medium access, queue management (\textit{v\_packets} being added, removed and discarded from buffers) and, eventually, the delivery of \textit{v\_packets} to their virtual destination nodes.
In particular, AFTER takes into account inter-flow interference, employing a set of deterministic rules to deal with nodes competing to access the wireless medium. However, AFTER cannot handle arbitrary traffic models because it seeks to estimate the maximum achievable network throughput by considering each flow to have an infinite backlog at the source. This means that AFTER provides no support for scenarios in which multimedia applications themselves limit the transmission rate of each flow. 

In summary, to our knowledge, MAPE is the first deterministic performance estimator that takes into account both inter-flow interference and heterogeneous flows, i.e., flows with different data rates, while being able to be executed in real-time.

\section{Multimedia-Aware Performance Estimator}\label{sec_MAPE}
As discussed in Section~\ref{sec_RelatedWorks}, although a number of performance estimation approaches have been proposed, none of them is able to provide real-time performance estimates that account for both inter-flow interference and rate-heterogeneous flows. The proposed Multimedia-Aware Performance Estimator, or MAPE, tries to fill this gap and uses a deterministic simulation-based approach to estimate the long term average throughput, packet loss and end-to-end delay for all (multi-rate) flows considering inter-flow interference. Note that considering multi-rate flows is essential to more realistically reproduce the behavior of multimedia applications.
For instance, in video applications, transmission rates are determined by video coding at each source and, therefore, each flow can be transmitted at different rates.

\subsection{MAPE - Design and Operation}
Algorithm~\ref{alg:MAPE} illustrates MAPE's overall operation, which is divided in three steps: \textbf{Step 1 -} MAPE starts with a complete snapshot of the current network state as input consisting of a representation of the network topology that includes link quality estimates (i.e., link frame delivery probability), list of currently active flows along with the respective paths, and each flow's data rate;  \textbf{Step 2 -} MAPE then uses the initial network snapshot to simulate the network as it operates until reaching steady state, which is used to compute long term throughput, packet loss, and end-to-end delay estimates in \textbf{Step 3}. Note that we employ the term \textit{steady state} in the same sense as in~\cite{passos2018after}, i.e., as a finite cycle of states that repeats themselves. 

\newlength{\textfloatsepsave} \setlength{\textfloatsepsave}{\textfloatsep} \setlength{\textfloatsep}{0pt}
\begin{algorithm}[!ht]
\algorithmiccomment{{\bf Step 1}: Initialization}\\
$networkTopology \leftarrow$graph representing the network\\
$flowsPath \leftarrow$ list of paths of all flows\\
$flowsRate \leftarrow$ list of bitrate of all flows\\
\algorithmiccomment{{\bf Step 2}: Simulation}\\
\While{no Steady State}{
\ForEach{flow $f \in flowsPath$}{
Update the number of \textit{v\_packets} received by flow $f$; \\
Schedule the queuing of new packet of flow $f$ to its queue according to $flowsRate$;\\
}
Next network state;
}
\algorithmiccomment{{\bf Step 3}: Estimation}\\
Compute long term per-flow performance. 
\caption{MAPE's pseudo-code.}
\label{alg:MAPE}
\end{algorithm}
\setlength{\textfloatsep}{\textfloatsepsave}

At the end of each iteration, MAPE stores a snapshot of the current network state, which consists of currently ongoing transmissions with their respective remaining times, the content of the queues and the backoff counter of the wireless medium access for all nodes that are traversed by any flow on the evaluated flow set, and the current medium access priority list. To decide whether the steady state has been achieved, the current state is compared to all previous ones. Whenever a duplicate state is found, MAPE declares that steady state has been reached and computes the average throughput, packet loss rate and end-to-end delay for each flow. 
A heuristic stop criterion is also used to guarantee low execution time and adequate real-time performance independent of application scenarios. When duplicated states are not found, MAPE computes the average cycle performance of events within which at least one packet from each flow has been delivered to its final destination as an attempt to approximate steady state performance.


Unlike stochastic simulators that study network behavior over a predefined period of time, MAPE aims at estimating the performance of the network, e.g., throughput, packet loss, and end-to-end delay at steady-state. 
This can be especially useful for QoS provisioning and, as previously noted, for route selection in real-time multimedia applications. Additionally, as discussed in Section~\ref{sec_RelatedWorks}, deterministic simulators that assume rate-homogeneous flows may result in severely inaccurate estimates for a number of reasons. First of all, the performance of a flow is necessarily limited by its transmission rate. Thus, such simulators may grossly overestimate performance in scenarios where network capacity is much larger than the aggregate demand of the active flows. Furthermore, severe underestimates may also occur for individual flows because interfering flows may be transmitted at a higher rate, consuming more network resources than they would in reality, reducing the achievable performance of other flows. MAPE overcomes these limitations by explicitly accounting for both multi-rate flows and inter-flow interference and thus attains more accurate  performance estimates in more realistic multimedia application scenarios. 

While MAPE builds on ``traditional'' deterministic estimators such as AFTER~\cite{passos2018after}, unlike these estimators, MAPE relaxes the assumption that all flows have infinite backlogs and instead generates \textit{v\_packets} according to the rate of  each flow --- which can be specified as an input, based on the flows' video coding rate, for instance. Whenever invoked, MAPE receives flow rate arguments as input and uses them deterministically to simulate the network dynamics by: (1) generating simulated \textit{v\_packets} for each flow at their respective source nodes, (2) triggering a number of other relevant simulation events, such as wireless medium access transmission, queue management (\textit{v\_packets} being added, removed and discarded from buffers), and, (3) eventually, delivering \textit{v\_packets} to their destination nodes. As such, inter-flow interference happens as a result of  buffer overflow, link-layer transmission losses, and medium access conflicts.


\subsection{MAPE - Implementation}
\begin{figure*}[!ht]
\centering
\includegraphics[width= 0.8 \linewidth]{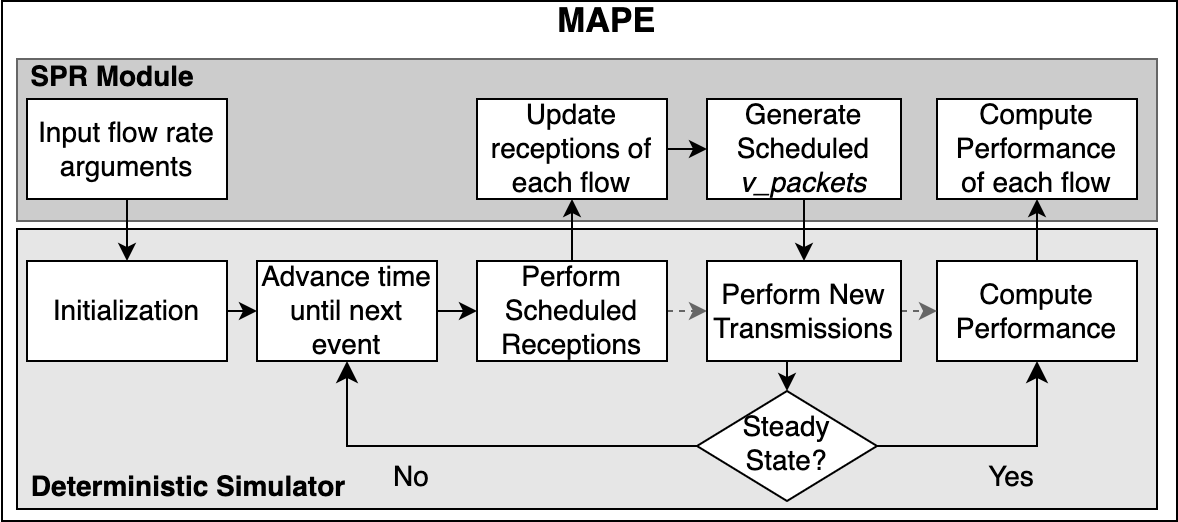}
\caption{MAPE's  implemention}
\label{fig:MFRmodule}
\end{figure*}

MAPE's current implementation uses AFTER~\cite{passos2018after} as the underlying deterministic performance estimator. As shown in Figure~\ref{fig:MFRmodule}, MAPE starts by initializing the simulation state with its input arguments. In this phase, the first packet of each flow is added to the queue of the respective source node and the simulation time is initialized to keep track of the events that are used to generate scheduled \textit{v\_packets}. Thus, the main loop of the simulation starts with the advance of the simulation according to the time of next possible events. This loop also handles packet receptions and eventually generates new transmission events until it detects that the network has reached a steady state --- a state when the same sequence of events starts to repeat itself --- which informs MAPE that it can then compute the estimated performance of each flow.

MAPE's functionality is implemented as a module (called SPR for Specific Per-flow Rates) that interfaces with the deterministic simulation engine to (1) provide flow rate information as part of simulation initialization, (2) update each flow when their \textit{v\_packets} are received, (3) generate new \textit{v\_packets} according to the stipulated flow rates, and (4) provide per-flow performance measurements. 

MAPE uses a representation of the current simulation state, which includes information about all received \textit{v\_packets}. Furthermore, the SPR module implements a procedure to schedule the next packet generation for each flow according to the specified rate and keeps track of the number of \textit{v\_packets} received per flow, which is used to calculate performance estimates for each flow. The number of received \textit{v\_packets} is also used to determine the steady state cycle -- i.e. the shortest sequence of network states that repeats itself indefinitely on the steady state of the simulation. 
While in the original AFTER the network state is updated when an  \textit{v\_packet} from any flow arrives at the destination, in the SPR module the condition was modified to update it only when all flows receive at least one \textit{v\_packet}.

To simulate \textit{v\_packet} transmissions, MAPE starts by placing the initial \textit{v\_packet} of each flow on the queue of the respective source node. It then iterates through all nodes that have at least one \textit{v\_packet} on their queues and triggers events for dequeuing a \textit{v\_packet} and adding this \textit{v\_packet} to a transmission buffer, where the \textit{v\_packet} is stored while waiting for an opportunity to be transmitted.

Note that MAPE's SPR Module introduces a mechanism to schedule the next \textit{v\_packet} generation for each flow according to the specified rate. Once per-flow rates have been specified, the scheduler uses them to place new \textit{v\_packets} in each source node's queue until the steady state is detected. Thus, \textit{v\_packets} of each flow are generated according to the intervals of the simulation time. To keep track of the simulated time, MAPE uses a time variable that is updated according to the end time of a link layer transmission attempt and the backoff procedure.

Once the simulation reaches steady state, MAPE computes the average throughput of each flow as the ratio between the total number of \textit{v\_packets} delivered within the last steady state cycle --- i.e., the period between two repeating simulation events --- the steady state cycle length.
In addition, the SPR Module 
computes packet loss and end-to-end delay by tracking 
all \textit{v\_packets} from the instant when they are generated at their source nodes until they are received at their destinations. MAPE is then able to estimate the average per-flow packet loss rate and end-to-end delay. Such metrics account for the data link layer transmission attempts and queuing delays.

\subsection{Discussion}

While MAPE makes  assumptions about network events and convergence to steady state, our experimental evaluation (see Section~\ref{sec_Results}) shows that MAPE is still able to estimate per-flow performance with adequate accuracy in quasi real time.  

Note that MAPE uses information about the topology of the network and the driving application (e.g., multimedia sources, flow rates), and, in the application scenarios envisioned (e.g., Smart Cities, Industrial Automation), nodes are typically stationary and have access to continuous power sources. As such, frequent topology changes (and energy limitations) are not expected to play a significant role. In scenarios where topology changes need to be considered, topology updates can be conveyed by proactive routing protocols. 

Route selection is an example of how MAPE can be used in practice. The routing protocol would invoke MAPE with an up-to-date network snapshot as input. Then, based on MAPE's performance estimates, it would perform route selection accordingly. For instance, a proactive link-state algorithm (e.g., OLSR~\cite{clausen2003}) can periodically discover topology changes and disseminate this information through link state  updates that MAPE can use to adjust its estimates. Network topology information would be updated whenever a node identifies “significant” changes in the network topology, e.g., link failures, new nodes/links or changes in link quality.
As part of our experimental evaluation (see Section~\ref{sec_Pratical}), we show how MAPE can guide route selection and, as a result, improve video transmission quality.

In its current implementation, MAPE assumes that flows are transmitted at constant bitrate. However, multimedia applications typically employ variable bitrate transmission. One way to address this is to simply have MAPE use the flow's average bitrate, which can be determined during transmission. Another approach to handle dynamic traffic patterns is to provide MAPE with updated data rate information whenever significant transmission rate changes are detected in the video coding process. In this work, we use the average bitrate of each video trace as input to MAPE. As part of future work, We plan to add support to variable bit rate flows.

\section{Evaluation Methodology}\label{sec_Experiments}

We evaluated MAPE against two types of discrete event simulators, stochastic (\textit{ns-3}) and deterministic (AFTER). We chose \textit{ns-3} because it is widely used by the network researchers and practitioners since it provides an adequate model of the network and, thus, provides reliable estimates of network performance. We use AFTER as the example of a deterministic simulator and demonstrate that MAPE can achieve better accuracy by being able to model specific per-flow rates, i.e., it simulates each flow transmitting at specified multimedia bitrates.

In this section, we describe the experimental methodology we use to evaluate MAPE, including the topologies and traffic models considered, as well as how the experiments were carried out.

\subsection{Experimental Topologies}
We evaluate MAPE using two different IoT wireless network topologies akin of IoT scenarios and whose parameters are summarized in Table~\ref{tab:paramScenarios}. The \textit{Random Indoor} topology aims to replicate smart building scenarios and was generated by placing nodes randomly within an indoor environment. The \textit{Grid Outdoor} topology tries to simulate smart city scenarios of neighborhood blocks and streets in an urban region, represented by a grid. More specifically, we reproduced a region of the city of Niterói, in the state of Rio de Janeiro, Brazil using an $8\times7$ grid of nodes in which two consecutive nodes are placed $60 m$ and $70 m$ apart on a given line and column of the grid, respectively.

\begin{table}[!h]
\small\sf\centering
\caption{Simulation scenarios}
\label{tab:paramScenarios}
\begin{tabular}{l|c|c}
\hline
\multicolumn{1}{c|}{\multirow{2}{*}{Parameters}} & \multicolumn{2}{c}{Topologies} \\ \cline{2-3} 
\multicolumn{1}{c|}{}                            & Random Indoor   & Grid Outdoor  \\ \hline
Deployment area ($m^2$):                             & $100$ x $100$       & $360$ x $490$     \\ 
Number of nodes:                                  & $30$              & $55$            \\ 
PHY/MAC technology:      & $802.11g$       & $802.11g$ \\
Link speed:      & $18$ $Mb/s$       & $18$ $Mb/s$  \\
Mac Queue Size:      & $10p$       & $10p$  \\
Packet Lifetime:   & $1000ms$       & $1000ms$  \\
Traffic Control Queue Size:      & $1p$       & $1p$  \\
Network Queue Size:      & $1p$       & $1p$  \\
Propagation Model:                                & Shadowing       & Cost231       \\
\hline
\end{tabular}
\end{table}

For a fair comparison between the two simulators, we set up the same link speeds, queue sizes and packet lifetime policy on \textit{ns-3} and MAPE.
The \textit{Shadowing} and \textit{Cost231} propagation models~\cite{Stoffers2012} were chosen to more realistically reproduce indoor and urban environments. All simulations use the same MAC and PHY technology and the same link speed, which was chosen to support multimedia application scenarios.

In order to estimate link quality (an information that is required by MAPE), a series of preliminary simulations were performed using the \textit{ns-3} simulator. For all nodes in each topology, we executed a simulation transmitting $20,000$ packets to extract the long term quality of each link.

\subsection{Traffic Models}\label{sec_Traffic}
In addition to link quality, MAPE requires per-flow transmission rate information. In our experiments, we use a mix of three different rates (as shown in Table~\ref{tab:trafficparameters}) to represent different levels of video quality. The EvalVid framework~\cite{klaue2003evalvid} was used to generate traces of the same video clip with these three rates, and the resulting average bitrate of each video trace was used as input to MAPE. Additional traffic generation parameters and their values used in our simulation experiments are listed in Table~\ref{tab:trafficparameters}. 

\begin{table}[!h]
\small\sf\centering
\caption{MM and CBR traffic parameters}
\label{tab:trafficparameters}
\begin{tabular}{l|l}
\hline
 \multicolumn{1}{c|}{Parameters}        & \multicolumn{1}{c}{Values}           \\ \hline
\multicolumn{2}{c}{MM-Traffic}\\\hline 
Video:            & Hall Monitor          \\ 
Encoding:         & H.264/MPEG-4 AVC \\ 
Frame rate:       & $30$ $Hz$                                 \\ 
Format:           & YUV CIF, $352$ x $288$                         \\ 
Number of frames: & $3600$                                   \\ 
Target Bitrate:          & $256$, $512$, $1024$ $kb/s$                   \\
Packet size:      & $1024$ $bytes$                 \\\hline
\multicolumn{2}{c}{CBR-Traffic}\\\hline 
Bitrate:      & $261$, $485$ and $836$ $kb/s$                 \\
Packet size:      & $1024$ $bytes$                 \\
\hline
\end{tabular}
\end{table}

Experiments which used multimedia (MM) traffic employ a publicly available and commonly used video clip, namely “Hall Monitor”~\cite{seeling2011}, which was converted to H.264 format with a rate of $30$ frames per second. Considering real-time transmission delay and human tolerance, the play-out buffer is set to $300ms$ to mitigate potential out-of-order packets; packets with delay longer than $300ms$ are discarded at the decoder.

In video traffic, transmission rates may vary according to the coding technique used. For example, more important video frames (e.g., MPEG I-frames) are often transmitted at higher rates than the target compression bitrate, while less important frames (e.g., MPEG P-frames and B-frames) are transmitted at lower rates. In our experiments, multimedia (MM) traffic target bitrates used by MAPE are based on long-term average bitrates calculated at the video source encoder. 
Because MAPE currently models variable bit rate flows using their long-term average rates, we also ran experiments with CBR traffic in our \textit{ns-3} simulations in order to assess how short-term fluctuations of the video traffic bitrate affect MAPE's estimates. 
In those experiments, we adopt the same bitrates used as input for AFTER and MAPE as listed in Table~\ref{tab:trafficparameters}. As part of our future work (see Section~\ref{sec_Conclusion}), we will modify MAPE's current variable bit rate traffic model to be able to account for shorter-term transmission rate variations.

\subsection{Experiments}
Simulation experiments were conducted as follows. For each topology, we computed the $5$ best paths (based on the quality of their links) for $500$ source-destination pairs generated randomly. Selecting one path for each pair, out of their $5$ best, we generated random instances for scenarios with $3$, $6$, $9$, and $12$ pairs (or flows), which are used to transmit concurrent video flows with $3$ different levels of quality -- a third of the flows use each of the three transmission rates listed in Table~\ref{tab:trafficparameters}. For instance, in a scenario with $6$ flows, we have $2$ sources transmitting at $256$ $kb/s$, $2$ sources transmitting at $512$ $kb/s$, and $2$ sources transmitting at $1024$ $kb/s$. 
We left out the evaluations of scenarios with more than $12$ flows because the networks become saturated. These scenarios do not provide satisfactory support for video applications, so they are not relevant for this work.

Finally, all scenarios were also executed in the \textit{ns-3} simulator for both the CBR and MM traffic models using a simulation time of $120 s$.
For each scenario, execution time, per-flow throughput, end-to-end delay, and packet loss were computed by averaging results over all runs.

\subsection{Evaluation Metrics}
We evaluate MAPE's performance according to execution time and prediction accuracy. Since \textit{ns-3} is a packet-level simulation platform, we are using its throughput, packet loss and end-to-end delay statistics as the ground truth in our performance study. 
Throughput is calculated as the ratio between the number of packets delivered to the destination and simulation time. End-to-end delay is the time interval between when a packet is transmitted by the source node and when that packet is delivered at the destination, averaged over all packets received, and packet loss is calculated as the percentage of packets transmitted that were not delivered to the destination. 

We expect MAPE to achieve predictions close to those of \textit{ns-3}, but in a reproducible manner and at a fraction of the required execution time. We also use SSIM~\cite{wang2004} and another metric called \textit{classification inversions} --- as defined in~\cite{passos2018after}, and further explained in Subsection~\ref{subsec:RoutingDecisions} --- to evaluate video quality and demonstrate the practical suitability of MAPE to the problem of route selection for multimedia applications.

\section{MAPE's Accuracy Evaluation}\label{sec_Results}
Our experimental evaluation aims at demonstrating MAPE's ability to accurately estimate per-flow throughput, delay and packet loss in a timely manner when compared to estimates provided by existing stochastic and  deterministic simulators. To this end, we compare MAPE against \textit{ns-3}~\cite{riley2010ns3} and AFTER~\cite{passos2018after} by considering the trade-off between  execution time and  throughput, delay and packet loss estimate accuracy. 

\subsection{Execution Time}
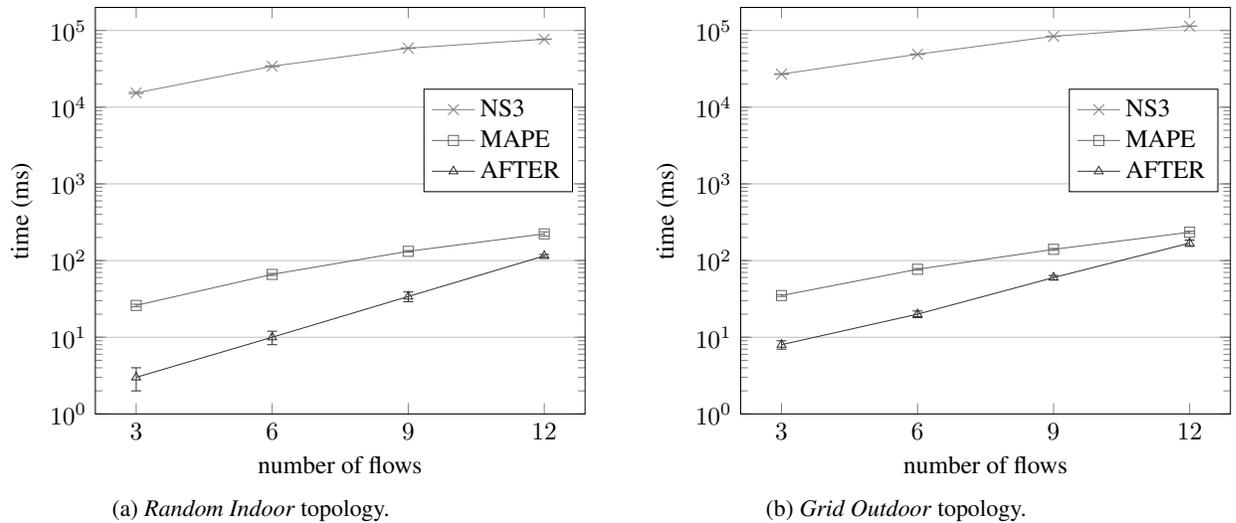
\begin{figure*}[!ht]
    \centering
    \begin{subfigure}[b]{0.4\linewidth}
    \begin{tikzpicture}[scale=0.95]
    
    \begin{axis}[ylabel={time (ms)}, xlabel={number of flows},xtick=data,
      legend pos=outer north east,
      legend cell align={left}, 
     ymin=1,
     ymax=200000,
     ymode=log,
     log basis y={10},
     ymajorgrids,
     legend style ={ at={(0.67,0.8)}, 
        anchor=north west, draw=black, 
        fill=white,align=left},
    cycle list name=black white,
    ]
    
    \addplot[color=gray,  mark size=3pt,mark=x, mark options={solid},error bars/.cd,y dir=both,y explicit] 
    coordinates {
            (3,15342) +- (300,300)
            (6,34188) +- (400,450)
            (9,58853) +- (450,450)
            (12,76929) +- (600,600)
        };
   
    \addlegendentry{NS3}  
    
    \addplot[color=black!60,mark=square, mark options={solid},error bars/.cd,y dir=both,y explicit] 
    coordinates {
            (3,26) +- (1,1)
            (6,66) +- (2,2)
            (9,132) +- (3,3)
            (12,223) +- (14,14)
        };
    \addlegendentry{MAPE}   
  
   \addplot[color=black!80,mark=triangle, mark options={solid},error bars/.cd,y dir=both,y explicit] 
    coordinates {
            (3,3) +- (1,1)
            (6,10) +- (2,2)
            (9,34) +- (3,5)
            (12,115) +- (5,5)
        };
    \addlegendentry{AFTER}  
         
    \end{axis}
    \end{tikzpicture}
    \caption{\textit{Random Indoor} topology.}
\label{fig:executionTime_random}
\end{subfigure}
\hspace{1.8cm}
 \begin{subfigure}[b]{0.4\linewidth}
 \begin{tikzpicture}[scale=0.95]
    
    \begin{axis}[ylabel={time (ms)}, xlabel={number of flows},xtick=data,
      legend pos= north east,
      ymin=1,
      ymax=200000,
     ymode=log,
     log basis y={10},
     ymajorgrids,
      legend style ={ at={(0.67,0.8)}, 
      legend cell align={left}, 
        anchor=north west, draw=black, 
        fill=white,align=left},
    cycle list name=black white,
    ]
    
    \addplot[color=gray,  mark size=3pt,mark=x, mark options={solid},error bars/.cd,y dir=both,y explicit] 
    coordinates {
            (3,27000) +- (300,300)
            (6,49000) +- (400,400)
            (9,84000) +- (450,450)
            (12,114000) +- (600,600)
        };
   
    \addlegendentry{NS3}   
    
    \addplot[color=black!60,mark=square, mark options={solid},error bars/.cd,y dir=both,y explicit] 
    coordinates {
            (3,35) +- (1,1)
            (6,77) +- (2,2)
            (9,140) +- (4,4)
            (12,234) +- (8,8)
        };
   
    \addlegendentry{MAPE}        
    
    \addplot[color=black!80,mark=triangle, mark options={solid},error bars/.cd,y dir=both,y explicit] 
    coordinates {
            (3,8) +- (1,1)
            (6,20) +- (2,2)
            (9,60) +- (4,4)
            (12,169) +- (16,16)
        };
   
    \addlegendentry{AFTER}

    \end{axis}
    \end{tikzpicture}
   \caption{\textit{Grid Outdoor} topology.}
\label{fig:executionTime_grid}
\end{subfigure}
\caption{Execution time (log scale) for different number of flows }
\label{fig:executionTime}
\end{figure*}

We measure average execution time for AFTER, MAPE, and \textit{ns-3} for each scenario considering the $95\%$ confidence intervals. All mean times are in milliseconds and simulations were performed on a dedicated server with an Intel i7-860 processor running at 2.8 GHz and 32 GB of RAM. As shown in Figure~\ref{fig:executionTime}, MAPE and AFTER report execution times that are at least $2$ orders of magnitude lower than those of \textit{ns-3} for different scenarios.

Note that execution times for \textit{ns-3} vary from tens to hundreds of seconds for the scenarios considered. While we observe a slight increase in MAPE's time complexity when compared to AFTER's for scenarios with only a few flows, that difference becomes negligible when the number of flows increases. It demonstrate that MAPE is able to compute per-flow network performance estimates in real time which can be used to inform core network services such as routing. MAPE and AFTER are fast because, unlike stochastic simulators, they do not need to simulate nearly as many events to reach steady state.

As expected, execution times increase with the number of flows. However, AFTER and MAPE's execution times increase more significantly with the number of path hops because that increases the number of transmission and reception events needed to deliver the flows'  \textit{v\_packets} to the destination node. This explains the slightly higher times measured with the \textit{Grid Outdoor} topology, which typically requires paths with more hops because of the greater distances between nodes. 

\subsection{Estimated Throughput}
\begin{figure*}[!ht]  
  \centering 
  \begin{subfigure}[b]{0.4\linewidth}
  \centering 
    \begin{tikzpicture}[scale=0.95]
\pgfplotsset{
    /pgfplots/bar cycle list/.style={/pgfplots/cycle list={
        {black!80,fill=black!80,mark=none},
        {gray,fill=gray,mark=none},
        {gray!60,fill=gray!60,mark=none},
        {black,fill=gray,mark=none},
}, },
}
    \begin{axis}[
        ybar,
        bar width=15pt,
        xtick distance=3,
        xlabel=number of flows,
        ylabel=ratio ($Prediction/Reference$),
        enlarge x limits={abs=1.5},
        ymin=0,
        ymax=10,
        xmin=3,
        xmax=12,
        ymajorgrids,
        scaled ticks=false,
        xtick style={
            /pgfplots/major tick length=0pt,
        },
        legend image code/.code={
            \draw[#1, draw=none] (0cm,-0.1cm) rectangle (0.3cm,0.2cm);
        },  
        legend pos= north east,
        legend cell align={left}, 
    ]
         \addplot+ [
            error bars/.cd,
                y dir=both,
                y explicit ,
        ] coordinates {
            (3,0.99) +- ( 0.05,  0.05)
            (6,1.15) +- (0.05,  0.05)
            (9,1.40) +- (0.05,  0.05)
            (12,1.48) +- (0.05,  0.05)
        };
        \addplot+ [
            error bars/.cd,
                y dir=both,
                y explicit,
        ] coordinates {
            (3,9.2) +- ( 0.5,  0.5)
            (6,5.16) +- (0.2,  0.2)
            (9,5.33) +- (0.1,  0.1)
            (12,6.52) +- (0.1,  0.3)
        };
         \addplot+ [draw=red, ultra thin,dashed,smooth] 
    coordinates {(0,1) ( 3,1) ( 6,1) ( 9,1) ( 12,1) ( 15,1) };
        
        \legend{
            MAPE,
            AFTER,
        }
    \end{axis}

\end{tikzpicture}
    \caption{CBR-Traffic} \label{fig:ratioGridCBR}  
  \end{subfigure}
  \hspace{1.5cm}
\begin{subfigure}[b]{0.4\linewidth}
\centering 
  \begin{tikzpicture}[scale=0.95]
\pgfplotsset{
    /pgfplots/bar cycle list/.style={/pgfplots/cycle list={
        {black!80,fill=black!80,mark=none},
        {gray,fill=gray,mark=none},
        {gray!60,fill=gray!60,mark=none},
        {black,fill=gray,mark=none},
}, },
}
    \begin{axis}[
        ybar,
        bar width=15pt,
        xtick distance=3,
        xmin=3,
        xmax=12,
        xlabel=number of flows,
        ylabel=ratio ($Prediction/Reference$),
        enlarge x limits={abs=1.5},
        ymin=0,
        ymax=10,
        ymajorgrids,
        scaled ticks=false,
        xtick style={
            /pgfplots/major tick length=0pt,
        },
        legend image code/.code={
            \draw[#1, draw=none] (0cm,-0.1cm) rectangle (0.3cm,0.2cm);
        },  
        legend pos= north east,
        legend cell align={left}, 
    ]
         \addplot+ [
            error bars/.cd,
                y dir=both,
                y explicit ,
        ] coordinates {
            (3,1.2) +- ( 0.01,  0.01)
            (6,1.3) +- (0.15,  0.15)
            (9,1.4) +- (0.01,  0.01)
            (12,1.6) +- (0.03,  0.03)
        };
        \addplot+ [
            error bars/.cd,
                y dir=both,
                y explicit,
        ] coordinates {
            (3,9.4) +- ( 0.4,  0.4)
            (6,5.7) +- (0.2,  0.2)
            (9,5.44) +- (0.1,  0.1)
            (12,6.03) +- (0.2,  0.2)
        };
        
         \addplot+ [draw=red, ultra thin,dashed,smooth] 
    coordinates {(0,1) ( 3,1) ( 6,1) ( 9,1) ( 12,1) ( 15,1) };

        \legend{
            MAPE,
            AFTER,
        }
    \end{axis}

\end{tikzpicture}
\caption{MM-Traffic} \label{fig:ratioGridMM}  
\end{subfigure}
\color{magenta}
\caption{Estimated throughput accuracy relative to \textit{ns-3}
in the Outdoor Grid topology.}\color{black}
\label{fig:ratio_grid}
\end{figure*}
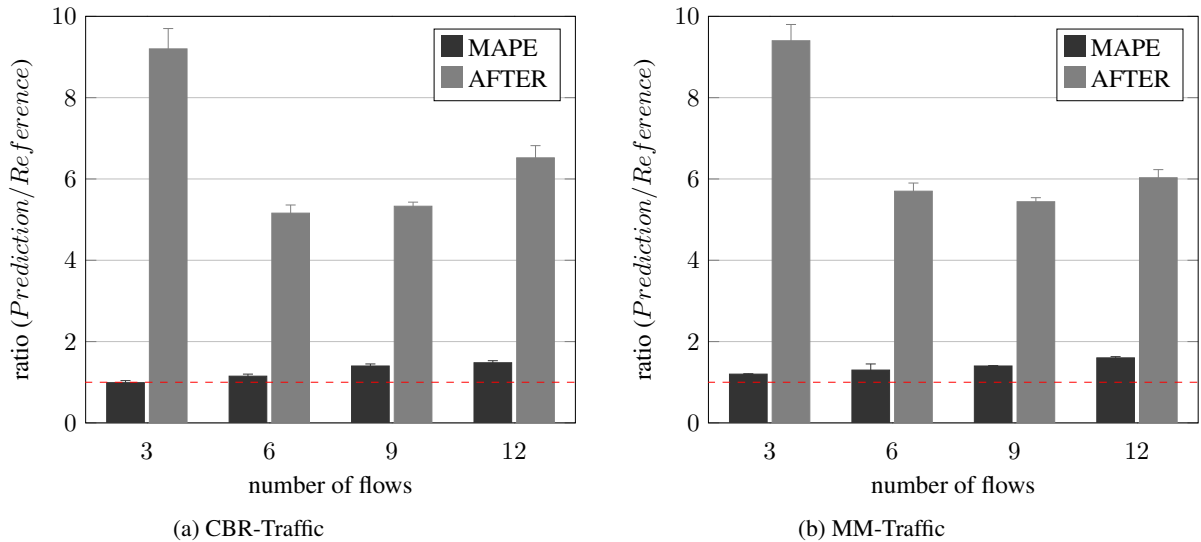

We measure throughput estimate accuracy as the ratio between the per-flow estimate returned by AFTER or MAPE and the per-flow throughput obtained by \textit{ns-3}. Differently from other common ways to measure accuracy, such as the mean squared error, the way we evaluate accuracy conveys whether the estimate is an underestimate or overestimate of the reference value, which is the value reported by \textit{ns-3}.  Figures~\ref{fig:ratio_random} and~\ref{fig:ratio_grid} show MAPE's and AFTER's throughput estimate accuracy for CBR and multimedia traffic in both the \textit{Random Indoor} and the \textit{Grid Outdoor} topologies, respectively. The red line represents the “ideal” ratio of $1$, i.e., a perfect match between the estimates and \textit{ns-3}'s measured throughput.

\begin{figure*}[!ht]  
  \centering 
  \begin{subfigure}[b]{0.4\linewidth}
  \centering 
    \begin{tikzpicture}[scale=0.95]
\pgfplotsset{
    /pgfplots/bar cycle list/.style={/pgfplots/cycle list={
        {black!80,fill=black!80,mark=none},
        {gray,fill=gray,mark=none},
        {gray!60,fill=gray!60,mark=none},
        {black,fill=gray,mark=none},
}, },
}
    \begin{axis}[
        ybar,
        bar width=15pt,
        xtick distance=3,
        xlabel=number of flows,
        ylabel=ratio ($Prediction/Reference$),
        enlarge x limits={abs=1.5},
        ymin=0,
        ymax=10,
        xmin=3,
        xmax=12,
        ymajorgrids,
        scaled ticks=false,
        xtick style={
            /pgfplots/major tick length=0pt,
        },
        legend image code/.code={
            \draw[#1, draw=none] (0cm,-0.1cm) rectangle (0.3cm,0.2cm);
        },  
        legend pos= north east,
    ]
         \addplot+ [
            error bars/.cd,
                y dir=both,
                y explicit ,
        ] coordinates {
            (3,1.03) +- ( 0.05,  0.05)
            (6,1.05) +- (0.05,  0.05)
            (9,1.23) +- (0.05,  0.05)
            (12,1.30) +- (0.05,  0.05)
        };
        \addplot+ [
            error bars/.cd,
                y dir=both,
                y explicit,
        ] coordinates {
            (3,7.8) +- ( 0.5,  0.5)
            (6,3.8) +- (0.2,  0.2)
            (9,2.9) +- (0.1,  0.1)
            (12,3.0) +- (0.1,  0.1)
        };

        \addplot+ [draw=red, ultra thin,dashed,smooth] 
    coordinates {(0,1) ( 3,1) ( 6,1) ( 9,1) ( 12,1) ( 15,1) };
        
        \legend{
            MAPE,
            AFTER,
        }
    \end{axis}

\end{tikzpicture}
    \caption{CBR-Traffic} \label{fig:ratioRandCBR}  
  \end{subfigure}
  \hspace{1.8cm}
\begin{subfigure}[b]{0.4\linewidth}
\centering 
  \begin{tikzpicture}[scale=0.95]
\pgfplotsset{
    /pgfplots/bar cycle list/.style={/pgfplots/cycle list={
        {black!80,fill=black!80,mark=none},
        {gray,fill=gray,mark=none},
        {gray!60,fill=gray!60,mark=none},
        {black,fill=gray,mark=none},
}, },
}
    \begin{axis}[
        ybar,
        bar width=15pt,
        xtick distance=3,
        xmin=3,
        xmax=12,
        xlabel=number of flows,
        ylabel=ratio ($Prediction/Reference$),
        enlarge x limits={abs=1.5},
        ymin=0,
        ymax=10,
        ymajorgrids,
        scaled ticks=false,
        xtick style={
            /pgfplots/major tick length=0pt,
        },
        legend image code/.code={
            \draw[#1, draw=none] (0cm,-0.1cm) rectangle (0.3cm,0.2cm);
        },  
        legend pos= north east,
    ]
         \addplot+ [
            error bars/.cd,
                y dir=both,
                y explicit ,
        ] coordinates {
            (3,1.05) +- ( 0.5,  0.05)
            (6,1.2) +- (0.5,  0.05)
            (9,1.5) +- (0.5,  0.05)
            (12,1.6) +- (0.5,  0.05)
        };
        \addplot+ [
            error bars/.cd,
                y dir=both,
                y explicit,
        ] coordinates {
            (3,8.6) +- ( 0.5,  0.5)
            (6,4.5) +- (0.2,  0.2)
            (9,3.5) +- (0.2,  0.1)
            (12,3.2) +- (0.2,  0.1)
        };
        
         \addplot+ [draw=red, ultra thin,dashed,smooth] 
    coordinates {(0,1) ( 3,1) ( 6,1) ( 9,1) ( 12,1) ( 15,1) };

        \legend{
            MAPE,
            AFTER,
        }
    \end{axis}

\end{tikzpicture}
\caption{MM-Traffic} \label{fig:ratioRandMM}  
\end{subfigure}
\color{magenta}
\caption{Estimated throughput accuracy relative to \textit{ns-3}
in Indoor Random topology.}\color{black}
\label{fig:ratio_random}
\end{figure*}
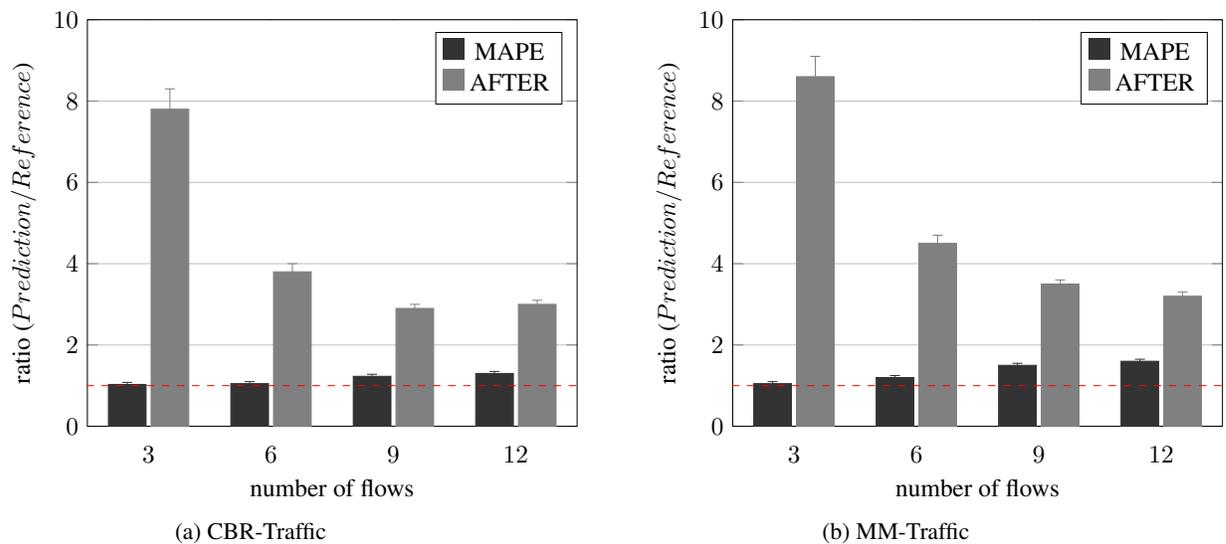 

We observe that AFTER's throughput estimates are significantly less accurate when compared to MAPE because AFTER's simulated flows attempt to transmit at the highest supported rate, typically resulting in overestimates. This is particularly pronounced for scenarios with few flows in which there is low inter-flow interference and, consequently, more residual network capacity to support higher transmission rates. As more flows are added, AFTER's prediction improves because, with more flows sharing the network's capacity, there is less room for each flow's transmission rate to increase beyond the real transmission rate.

This prediction discrepancy between AFTER and MAPE also quantitatively demonstrates the impact that not accounting for specific flow transmission rates may have. It also illustrates that MAPE is able to significantly improve prediction accuracy for scenarios with few flows (in our experiments, $3$- and $6$-flow scenarios). MAPE's accuracy decreases in scenarios with more flows --- with a bias toward overestimates due to some simplifications inherited from AFTER. For instance, AFTER does not take into account packet losses due to collision, which may influence network throughput when there are more flows transmitting simultaneously. Instead, in its inter-flow interference model, AFTER implements a medium access scheduler based on an interference graph of the topology. In future work, we plan to address this issue by improving how flow interference is modeled.

Note that MAPE yields higher accuracy for CBR traffic (Figures~\ref{fig:ratioRandCBR} and ~\ref{fig:ratioGridCBR}). That is because its scheduler also generates \textit{v\_packets} at constant rates. For multimedia (MM) traffic scenarios (Figures~\ref{fig:ratioRandMM} and ~\ref{fig:ratioGridMM}), however, transmission rate variations cause MAPE to overestimate the throughput. This is because bursts of the more important video packets cause losses due to buffer overflow and packet collisions, while less important video packets which have lower transmission rates are delivered more reliably. 

\begin{figure*}[!ht]  
  \centering 
  \begin{subfigure}[b]{0.4\linewidth}
  \centering 
    \begin{tikzpicture}[scale=0.95]
\pgfplotsset{
    /pgfplots/bar cycle list/.style={/pgfplots/cycle list={
        {black!60,fill=black!60,mark=none},
        {black!80,fill=black!80,mark=none},
        {gray,fill=gray,mark=none},
        {gray!60,fill=gray!60,mark=none},
}, },
}
    \begin{axis}[
        ybar,
        bar width=8pt,
        xtick ={1,3,5},
        xticklabels={261kb/s,485kb/s,836kb/s},
        xlabel=flow rates,
        ylabel=throughput (kb/s),
        enlarge x limits={abs=1.5},
        ymin=1,
        ymax=2200,
        xmin=0.5,
        xmax=5,
        ymajorgrids,
        scaled ticks=false,
        xtick style={
            /pgfplots/major tick length=0pt,
        },
        legend image code/.code={
            \draw[#1, draw=none] (0cm,-0.1cm) rectangle (0.3cm,0.2cm);
        },  
        legend pos= north west,
        legend cell align={left}, 
    ]
         
         \addplot+ [
            error bars/.cd,
                y dir=both,
                y explicit ,
        ] coordinates {
            (1,260) +- ( 1,  1)
            (3,477) +- (3,  3)
            (5,691) +- (12,  12)
        };
        \addplot+ [
            error bars/.cd,
                y dir=both,
                y explicit,
        ] coordinates {
            (1,229) +- ( 1,  1)
            (3,412) +- (1,  1)
            (5,640) +- (3,  3)
        };
        \addplot+ [
            error bars/.cd,
                y dir=both,
                y explicit,
        ] coordinates {
            (1,233) +- ( 2,  2)
            (3,370) +- (3,  3)
            (5,654) +- (5,  5)
        };
        \addplot+ [
            error bars/.cd,
                y dir=both,
                y explicit ,
        ] coordinates {
            (1,1600) +- ( 98,  98)
            (3,1585) +- (83,  83)
            (5,1634) +- (87,  87)
        };
        
        \addplot+ [draw=red, ultra thin,dashed,smooth] 
    coordinates {(0,261) ( 2,261)};
        \addplot+ [draw=red, ultra thin,dashed,smooth] 
    coordinates {(2,485) ( 4,485)};
        \addplot+ [draw=red, ultra thin,dashed,smooth] 
    coordinates {(4,836) ( 6,836)};

        \legend{
            MAPE,
            NS3-CBR,
            NS3-MM,
            AFTER
        }
    \end{axis}

\end{tikzpicture}
    \caption{\textit{Random Indoor} Topology.} \label{fig:throughput6flow-Rand}  
  \end{subfigure}
  \hspace{1.8cm}
\begin{subfigure}[b]{0.4\linewidth}
\centering 
   \begin{tikzpicture}[scale=0.95]
\pgfplotsset{
    /pgfplots/bar cycle list/.style={/pgfplots/cycle list={
        {black!60,fill=black!60,mark=none},
        {black!80,fill=black!80,mark=none},
        {gray,fill=gray,mark=none},
        {gray!60,fill=gray!60,mark=none},
}, },
}
    \begin{axis}[
        ybar,
        bar width=8pt,
        xtick ={1,3,5},
        xticklabels={261kb/s,485kb/s,836kb/s},
        xlabel=flow rates,
        ylabel=throughput (kb/s),
        enlarge x limits={abs=1.5},
        ymin=1,
        ymax=2200,
        xmin=0.5,
        xmax=5,
        ymajorgrids,
        scaled ticks=false,
        xtick style={
            /pgfplots/major tick length=0pt,
        },
        legend image code/.code={
            \draw[#1, draw=none] (0cm,-0.1cm) rectangle (0.3cm,0.2cm);
        },  
        legend pos= north west,
        legend cell align={left}, 
    ]
        
         \addplot+ [
            error bars/.cd,
                y dir=both,
                y explicit ,
        ] coordinates {
            (1,248) +- (3,  3)
            (3,377) +- (11,  11)
            (5,597) +- (20,  20)
        };
        \addplot+ [
            error bars/.cd,
                y dir=both,
                y explicit,
        ] coordinates {
            (1,201) +- ( 2,  2)
            (3,335) +- (5,  5)
            (5,487) +- (10,  10)
        };
        \addplot+ [
            error bars/.cd,
                y dir=both,
                y explicit,
        ] coordinates {
            (1,231) +- ( 2.5,  2.5)
            (3,356) +- (4.1,  4.1)
            (5,492) +- (6.7,  6.7)
        };
         \addplot+ [
            error bars/.cd,
                y dir=both,
                y explicit ,
        ] coordinates {
            (1,1963) +- (93,  93)
            (3,1900) +- (68,  68)
            (5,1882) +- (60,  60)
        };
        
        \addplot+ [draw=red, ultra thin,dashed,smooth] 
    coordinates {(0,261) ( 2,261)};
        \addplot+ [draw=red, ultra thin,dashed,smooth] 
    coordinates {(2,485) ( 4,485)};
        \addplot+ [draw=red, ultra thin,dashed,smooth] 
    coordinates {(4,836) ( 6,836)};

        \legend{
            MAPE,
            NS3-CBR,
            NS3-MM,
            AFTER
        }
    \end{axis}
\end{tikzpicture}
\caption{\textit{Grid Outdoor} Topology.} \label{fig:throughput6flow_Grid}  
\end{subfigure}
\caption{Average throughput for scenarios with 6 flows.}
\label{fig:throughput6flows}
\end{figure*}
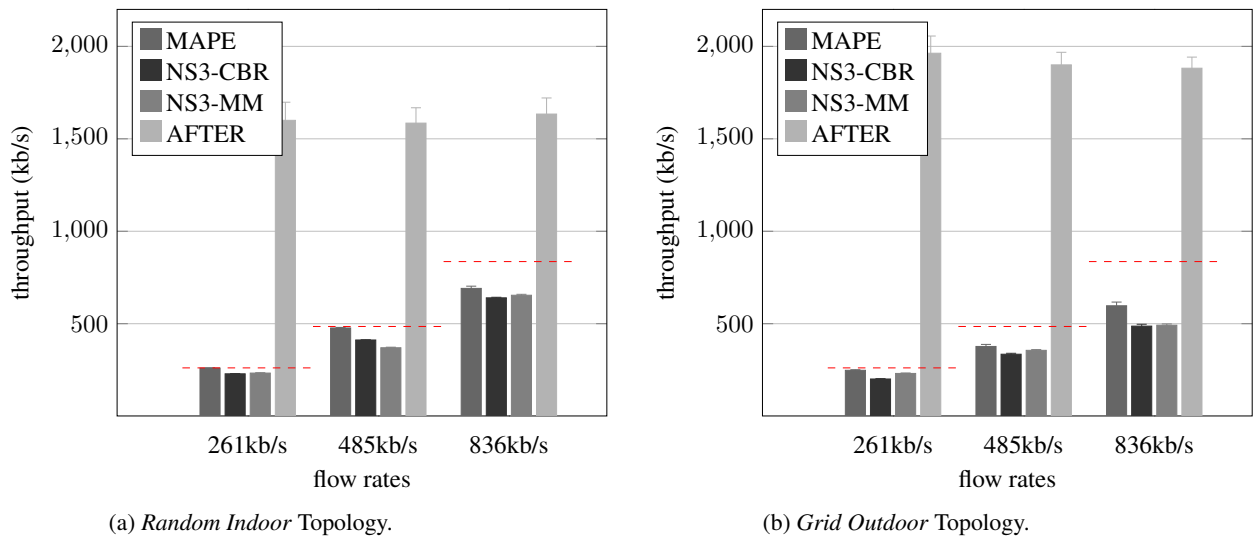

We also evaluate the per-flow throughput prediction accuracy considering the different classes of flows based on their transmission rates. 
Figure~\ref{fig:throughput6flows} summarizes the results for $6$ flows using CBR and MM traffic in both the indoor and outdoor topologies. We also ran these experiments for $3$, $9$ and $12$ flows, but we omit those results since they show similar trends. The red reference lines represent the ideal throughput based on the average bitrate generated for each video trace.

The figure shows that AFTER tends to overestimate all three classes of flows. Moreover, as the source-destination pair is chosen randomly regardless of the transmission rate of the flow, the average throughput estimated by AFTER tends to be roughly the same for all three classes. Conversely, by knowing the transmission rate of each flow, MAPE is able to more accurately estimate per-flow throughput. Note, however, that it slightly overestimates MM's throughput. This issue, which is more pronounced in the outdoor topology due to its higher link reliability, is due to the fact that MAPE's current implementation uses AFTER, and thus it inherits the mechanism used by AFTER to estimate packet loss. It considers two possible sources of packet loss: buffer overflow and link-layer transmission losses. If all links that compose a path have perfect delivery rates, losses computed by AFTER are only due to buffer overflow. In practice, however, there are other sources of losses, such as collisions, and as a result, MAPE and AFTER tend to overestimate flows' throughputs.

\subsection{Estimated Delay and Packet Loss }

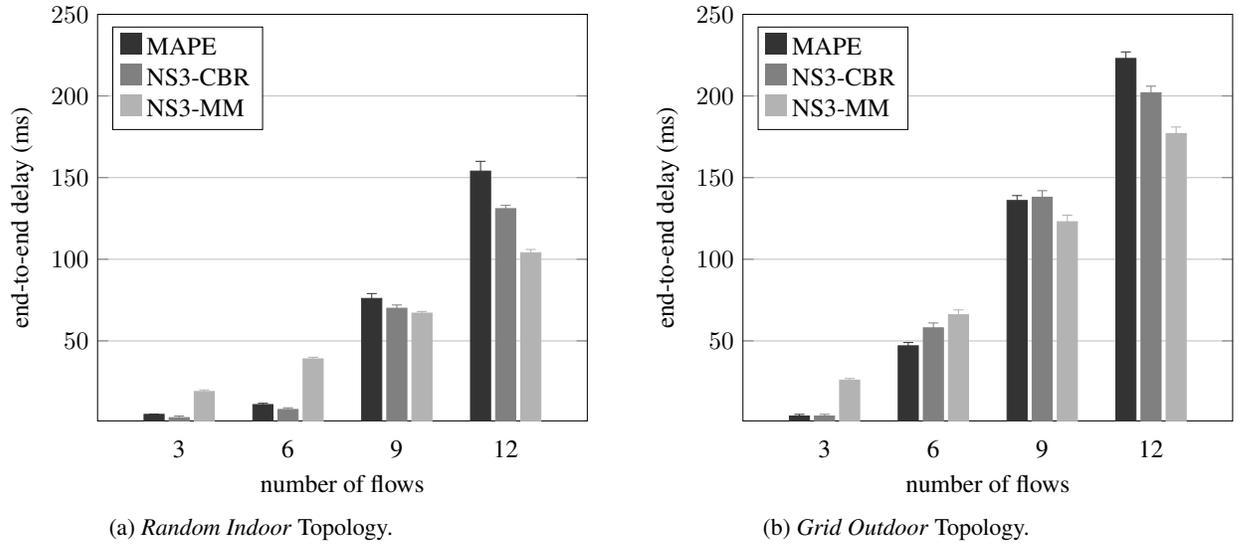
\begin{figure*}[!ht]  
  \centering 
  \begin{subfigure}[b]{0.4\linewidth}
  \centering 
    \begin{tikzpicture}[scale=0.95]
\pgfplotsset{
    /pgfplots/bar cycle list/.style={/pgfplots/cycle list={
        {black!80,fill=black!80,mark=none},
        {gray,fill=gray,mark=none},
        {gray!60,fill=gray!60,mark=none},
}, },
}
    \begin{axis}[
        ybar,
        bar width=8pt,
        xtick ={0,2,4,6},
        xticklabels={3,6,9,12},
        xlabel=number of flows,
        ylabel=end-to-end delay (ms),
        enlarge x limits={abs=1.5},
        ymin=1,
        ymax=250,
        xmin=0,
        xmax=6,
        ymajorgrids,
        scaled ticks=false,
        xtick style={
            /pgfplots/major tick length=0pt,
        },
        legend image code/.code={
            \draw[#1, draw=none] (0cm,-0.1cm) rectangle (0.3cm,0.2cm);
        },  
        legend pos= north west,
        legend cell align={left}, 
    ]
         \addplot+ [
            error bars/.cd,
                y dir=both,
                y explicit ,
        ] coordinates {
            (0,5) +- ( 0.1,  0.1)
            (2,11) +- (0.8,  0.8)
            (4,76) +- (3,  3)
            (6,154) +- (6,6)
        };
         \addplot+ [
            error bars/.cd,
                y dir=both,
                y explicit ,
        ] coordinates {
            (0,3) +- ( 1,  1)
            (2,8) +- (1,  1)
            (4,70) +- (2,  2)
            (6,131) +- (2,  2)
        };
        \addplot+ [
            error bars/.cd,
                y dir=both,
                y explicit,
        ] coordinates {
            (0,19) +- ( 1,  1)
            (2,39) +- (1,  1)
            (4,67) +- (1,  1)
            (6,104) +- (2,  2)
        };
        \legend{
            MAPE,
            NS3-CBR,
            NS3-MM
        }
    \end{axis}

\end{tikzpicture}
    \caption{\textit{Random Indoor} Topology.} \label{fig:delay-Rand}  
  \end{subfigure}
  \hspace{1.8cm}
\begin{subfigure}[b]{0.4\linewidth}
\centering 
   \begin{tikzpicture}[scale=0.95]
\pgfplotsset{
    /pgfplots/bar cycle list/.style={/pgfplots/cycle list={
        {black!80,fill=black!80,mark=none},
        {gray,fill=gray,mark=none},
        {gray!60,fill=gray!60,mark=none},
}, },
}
    \begin{axis}[
        ybar,
        bar width=8pt,
        xtick ={0,2,4,6},
        xticklabels={3,6,9,12},
        xlabel=number of flows,
        ylabel=end-to-end delay (ms),
        enlarge x limits={abs=1.5},
        ymin=1,
        ymax=250,
        xmin=0,
        xmax=6,
        ymajorgrids,
        scaled ticks=false,
        xtick style={
            /pgfplots/major tick length=0pt,
        },
        legend image code/.code={
            \draw[#1, draw=none] (0cm,-0.1cm) rectangle (0.3cm,0.2cm);
        },  
        legend pos= north west,
        legend cell align={left}, 
    ]
         \addplot+ [
            error bars/.cd,
                y dir=both,
                y explicit ,
        ] coordinates {
            (0,4) +- ( 1,  1)
            (2,47) +- (2,2)
            (4,136) +- (3,  3)
            (6,223) +- (4,4)
        };
         \addplot+ [
            error bars/.cd,
                y dir=both,
                y explicit ,
        ] coordinates {
            (0,4) +- ( 1,  1)
            (2,58) +- (3,  3)
            (4,138) +- (4,  4)
            (6,202) +- (4,4)
        };
        \addplot+ [
            error bars/.cd,
                y dir=both,
                y explicit,
        ] coordinates {
            (0,26) +- ( 1,  1)
            (2,66) +- (3,  3)
            (4,123) +- (4,  4)
            (6,177) +- (4,4)
        };
        \legend{
            MAPE,
            NS3-CBR,
            NS3-MM
        }
    \end{axis}

\end{tikzpicture}
\caption{\textit{Grid Outdoor} Topology.} \label{fig:delay-Grid}  
\end{subfigure}
\caption{Average end-to-end delay for different number of flows.}
\label{fig:delay}
\end{figure*}

We also evaluate MAPE's delay and packet loss estimates. Figure~\ref{fig:delay} shows the average end-to-end delay, considering the $95\%$ confidence intervals for different number of  flows in both experimental topologies. When compared to the results obtained by \textit{ns-3}, MAPE shows similar delay increase trend as the number of flows increases. 
Note that MAPE overestimates the end-to-end delay for scenarios with 12 flows in both topologies. This is due to MAC layer congestion as more flows share the same nodes/links increasing contention and consequently increasing MAPE's time to reach steady state, which, in turn, may cause MAPE's execution to end before reaching steady state.
Although MAPE's estimate is less accurate compared to \textit{ns-3} when it does not reach steady state, we will demonstrate in the Section~\ref{sec_Pratical} that these results are still useful to inform the route selection process ahead of multimedia flow transmission.

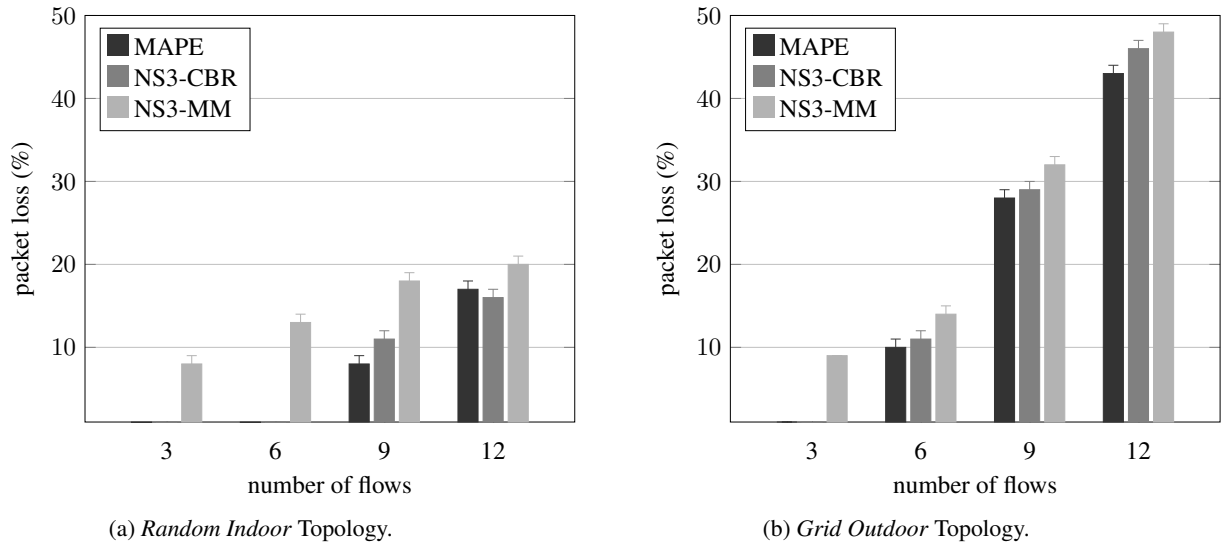
\begin{figure*}[!ht]  
  \centering 
  \begin{subfigure}[b]{0.4\linewidth}
  \centering 
    \begin{tikzpicture}[scale=0.95]
\pgfplotsset{
    /pgfplots/bar cycle list/.style={/pgfplots/cycle list={
        {black!80,fill=black!80,mark=none},
        {gray,fill=gray,mark=none},
        {gray!60,fill=gray!60,mark=none},
}, },
}
    \begin{axis}[
        ybar,
        bar width=8pt,
        xtick ={0,2,4,6},
        xticklabels={3,6,9,12},
        xlabel=number of flows,
        ylabel=packet loss (\%),
        enlarge x limits={abs=1.5},
        ymin=1,
        xmin=0,
        xmax=6,
        ymajorgrids,
        ymax=50,
        scaled ticks=false,
        xtick style={
            /pgfplots/major tick length=0pt,
        },
        legend image code/.code={
            \draw[#1, draw=none] (0cm,-0.1cm) rectangle (0.3cm,0.2cm);
        },  
        legend pos= north west,
        legend cell align={left}, 
    ]
         \addplot+ [
            error bars/.cd,
                y dir=both,
                y explicit ,
        ] coordinates {
            (0,0.05) +- ( 0,  0)
            (2,0.4) +- (0.2,  0.2)
            (4,8) +- (1,  1)
            (6,17) +- (1,1)
        };
         \addplot+ [
            error bars/.cd,
                y dir=both,
                y explicit ,
        ] coordinates {
            (0,0) +- ( 0,  0)
            (2,0.8) +- (0.2,  0.2)
            (4,11) +- (1,  1)
            (6,16) +- (1,1)
        };
        \addplot+ [
            error bars/.cd,
                y dir=both,
                y explicit,
        ] coordinates {
            (0,8) +- ( 1,  1)
            (2,13) +- (1,  1)
            (4,18) +- (1,  1) 
            (6,20) +- (1,  1)
        };
        \legend{
            MAPE,
            NS3-CBR,
            NS3-MM
        }
    \end{axis}

\end{tikzpicture}
    \caption{\textit{Random Indoor} Topology.} \label{fig:loss-RAnd}  
  \end{subfigure}
  \hspace{1.8cm}
\begin{subfigure}[b]{0.4\linewidth}
\centering 
   \begin{tikzpicture}[scale=0.95]
\pgfplotsset{
    /pgfplots/bar cycle list/.style={/pgfplots/cycle list={
        {black!80,fill=black!80,mark=none},
        {gray!,fill=gray!,mark=none},
        {gray!60,fill=gray!60,mark=none},
}, },
}
    \begin{axis}[
        ybar,
        bar width=8pt,
        xtick ={0,2,4,6},
        xticklabels={3,6,9,12},
        xlabel=number of flows,
        ylabel=packet loss (\%),
        enlarge x limits={abs=1.5},
        ymin=1,
        ymax=50,
        xmin=0,
        xmax=6,
        ymajorgrids,
        scaled ticks=false,
        xtick style={
            /pgfplots/major tick length=0pt,
        },
        legend image code/.code={
            \draw[#1, draw=none] (0cm,-0.1cm) rectangle (0.3cm,0.2cm);
        },  
        legend pos=north west,
        legend cell align={left}, 
    ]
         \addplot+ [
            error bars/.cd,
                y dir=both,
                y explicit ,
        ] coordinates {
            (0,1) +- ( 0,  0)
            (2,10) +- (1,1)
            (4,28) +- (1,  1)
            (6,43) +- (1,1)
        };
         \addplot+ [
            error bars/.cd,
                y dir=both,
                y explicit ,
        ] coordinates {
            (0,0) +- ( 0,  0)
            (2,11) +- (1,  1)
            (4,29) +- (1,  1)
            (6,46) +- (1,1)
        };
        \addplot+ [
            error bars/.cd,
                y dir=both,
                y explicit,
        ] coordinates {
            (0,9) +- ( 0,  0)
            (2,14) +- (1,  1)
            (4,32) +- (1,  1)
            (6,48) +- (1,1)
        };
        \legend{
            MAPE,
            NS3-CBR,
            NS3-MM
        }
    \end{axis}

\end{tikzpicture}
\caption{\textit{Grid Outdoor} Topology.} \label{fig:loss-Grid}  
\end{subfigure}
\caption{Average packet loss for different number of flows.}
\label{fig:packetloss}
\end{figure*}

Figure~\ref{fig:packetloss} plots the average packet loss rate. It also shows a discrepancy between MAPE's and \textit{ns-3}'s estimates in both topologies. But here, instead of overestimating, losses are generally underestimated by MAPE. The culprit is the absence of a collision packet loss counter in MAPE, which causes it to be more prone to estimate lower overall loss rates. These results also help explain the reason for instances in which MAPE overestimates the throughput ---  a consequence of fewer packets being discarded at the MAC layer. Furthermore, as expected, packet losses for MM traffic were even more impacted by the bursty nature of the video packet flows. As part of our future work, we plan to improve how MAPE models packet losses due to collision.

Despite those discrepancies, the results shown in Figures~\ref{fig:throughput6flows}, ~\ref{fig:delay} and~\ref{fig:packetloss} demonstrate MAPE's ability to capture the overall trend in throughput, delay and packet loss for multimedia flows in different application scenarios. Furthermore, we note that the discrepancies for $9$ and $12$ flows are mostly caused by network congestion and MAPE estimates being generated before steady state is achieved.

\begin{figure}[!ht]  
    \centering
  \begin{tikzpicture}[scale=0.85]
  \pgfplotsset{ scale only axis}
    \begin{axis}[
    xlabel=instances,
    ylabel=delay discrepancy ($Prediction-Reference$) ms,
    xmin=0,
    xmax=500,
    ymin=-100,
    ymax=460,
    legend pos=north east,
    label style = {font = {\fontsize{11 pt}{11 pt}\selectfont}},
    legend style = {font = {\fontsize{12 pt}{12 pt}\selectfont}},
    legend cell align={left}
    ]
    \addplot[
        only marks,
        mark color=black!80,
        mark=*]
    table[meta=delay]
    {data/gapDelaySteady.dat};
    \addplot[
        only marks,
        mark color=gray!80,
        mark=+]
    table[meta=delay]
    {data/gapDelayNonSteady.dat};
    \legend{
        reached steady state,
        did not reach steady state
    }
    \end{axis}
\end{tikzpicture}
    \caption{Difference between MAPE's and \textit{ns-3}'s average delay predictions considering MAPE's steady and non-steady instances for scenarios with 9 flows in the \textit{Random Indoor} Topology.}
    \label{fig:gapDelay}
\end{figure}
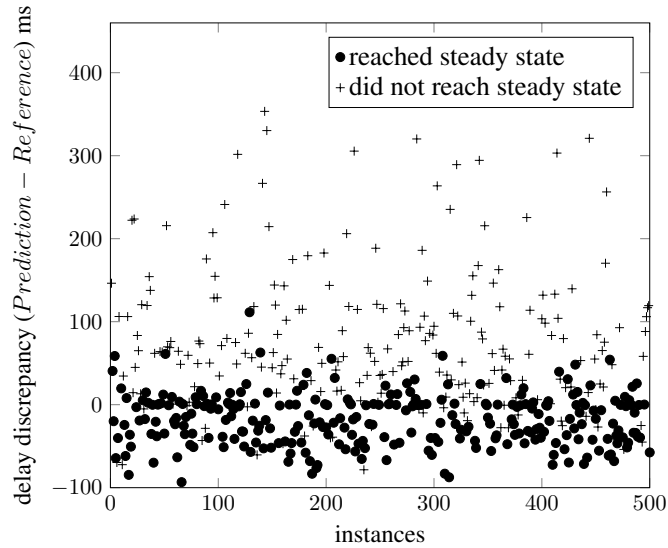

In order to confirm this hypothesis, in Figure~\ref{fig:gapDelay} we show a scatter plot for the $9$-flow runs using the \textit{Random Indoor} topology representing which instances did and did not reach the steady state and their respective delays discrepancies when comparing MAPE to \textit{ns-3} --- i.e., the difference between MAPE's and \textit{ns-3}'s average delay estimates.
Note that we show results for the $9$-flow \textit{Random Indoor} topology experiments because, with 9 flows (and above), the network gets more congested and consequently the number of instances that do not reach steady state increases, which, as previously discussed, results in higher delay and packet loss discrepancies.

As the plot shows, when steady state is reached, MAPE yields adequate estimation accuracy, with discrepancies concentrating around less than $100ms$. However, MAPE tends to overestimate end-to-end delay for instances that do not reach steady state, causing higher discrepancies.

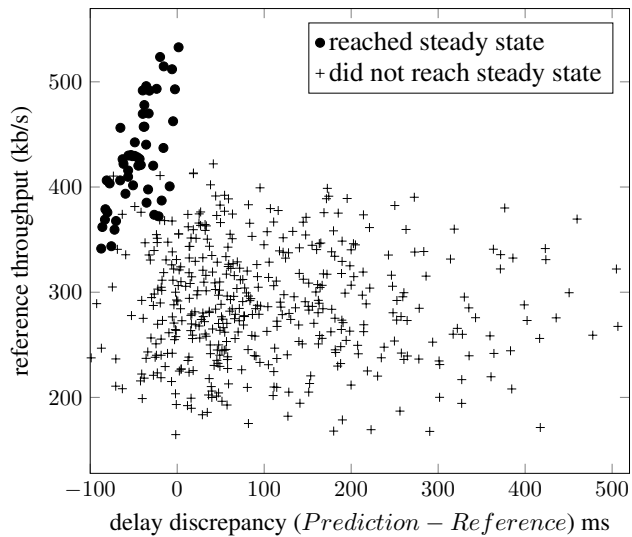
\begin{figure}[!ht]  
    \centering
  \begin{tikzpicture}[scale=0.85]
  \pgfplotsset{ scale only axis}
    \begin{axis}[
    xlabel=delay discrepancy ($Prediction-Reference$) ms,
    ylabel=reference throughput (kb/s),
    label style = {font = {\fontsize{11 pt}{11 pt}\selectfont}},
    xmin=-100,
    xmax=520,
    legend pos=north east,
    legend style = {font = {\fontsize{12 pt}{12 pt}\selectfont}},
    legend cell align={left}
    ]
    \addplot[
        only marks,
        mark color=black!80,
        mark=*]
    table[meta=delay]
    {data/diffDelaySteady.dat};
    \addplot[
        only marks,
        mark color=black!80,
        mark=+]
    table[meta=delay]
    {data/diffDelayNonSteady.dat};
   \legend{
        reached steady state,
        did not reach steady state,
    }

    \end{axis}
     
\end{tikzpicture}
    \caption{Path average throughput (according to \textit{ns-3}) as a function of the difference between MAPE's and \textit{ns-3}'s average delay predictions for scenarios with 12 flows in the \textit{Random Indoor} Topology.}
    \label{fig:diffDelay}
\end{figure}

Figure~\ref{fig:diffDelay} confirms this observation --- it shows a scatter plot of the path average throughput according to \textit{ns-3} for instances that reached the steady state and those that did not as a function of the delay estimate discrepancies for the $12$-flow experiments in the \textit{Random Indoor} Topology (since they showed the highest discrepancies). As the plot shows, higher throughput paths are concentrated around the smallest discrepancies, while the largest discrepancies happen with paths that present poor network performance and are, thus, not suitable for multimedia flows. Note that instances with higher throughput were those in which MAPE was able to reach steady state, unlike runs that exhibit higher discrepancies, which, again, are the ones where steady state was not reached.

As a tool to guide
real-time route selection decisions for IoT multimedia applications, low throughput routes ---likely because of congestion --- are generally undesirable, as they are often unable to meet the requirements of multimedia flows. As such, overestimating delay for those paths should not negatively impact path selection. That is, MAPE’s delay overestimates when compared to \textit{ns-3}'s correspond to paths that are undesirable for video traffic anyway and therefore should not be selected by routing.

\section{Video Quality Evaluation}\label{sec_Pratical}

The ability to estimate network performance is essential to ensure adequate network support for many IoT multimedia applications. In the case of applications involving video transmission, for instance, timely and fresh estimates of the current state of the network can significantly help routing protocols to rapidly identify paths that satisfy QoS constraints, as well as promote load balancing and network resource utilization. 
To examine how MAPE's performance estimates can be used to improve overall video quality, we use a well-known Quality-of-Experience (QoE) metric called Structural Similarity Index Measure (SSIM)~\cite{wang2004} measured by the EvalVid video transmission and quality evaluation framework~\cite{klaue2003evalvid}. In the second part of this section, we evaluate the quality of the video transmitted using the route selected based on MAPE's estimates.

\subsection{Video Structural Similarity}

The Structural Similarity Index Measure, or SSIM, measures video structural distortion which is known to correlate with video quality as perceived by the end user~\cite{wang2004}. This metric combines luminance, contrast, and structural similarity of the frames to compute the correlation between the original frame and the (possibly distorted) displayed one.
SSIM values vary between $0$ and $1$, with higher values meaning better quality.

To show how MAPE estimates can be used to improve  video quality, we run experiments transmitting the "Hall Monitor" video clip (as described in Section~\ref{sec_Traffic})
and compute the SSIM by comparing all transmitted and received video frames. 

\begin{figure}[!ht]
\centering 
   \begin{tikzpicture}[scale=1.05]
\pgfplotsset{
    /pgfplots/bar cycle list/.style={/pgfplots/cycle list={
        {black!80,fill=black!80,mark=none},
        {gray!,fill=gray!,mark=none},
        {gray!60,fill=gray!60,mark=none},
}, },
}
    \begin{axis}[
        ybar,
        bar width=14pt,
        xtick ={0,1,2,3},
        xticklabels={3,6,9,12},
        xlabel=number of flows,
        ylabel=SSIM,
        enlarge x limits={abs=0.3},
        ymin=0.5,
        ymax=1,
        xmin=0,
        xmax=3.3,
        ymajorgrids,
        scaled ticks=false,
        xtick style={
            /pgfplots/major tick length=0pt,
        },
        legend image code/.code={
            \draw[#1, draw=none] (0cm,-0.1cm) rectangle (0.3cm,0.2cm);
        },
        legend pos= south west,
        legend cell align={left}, 
    ]
       
         \addplot+ [
            error bars/.cd,
                y dir=both,
                y explicit ,
        ] coordinates {
            (1,0) +- ( 0,  0)
            (1,0.92) +- (0.01,0.01)
            (2,0.91) +- (0.01,  0.01)
            (3,0.86) +- (0.01,0.01)
        };
        
        \addplot+ [
            error bars/.cd,
                y dir=both,
                y explicit,
        ] coordinates {
            (0,0.99) +- ( 0,  0)
            (1,0.98) +- (0.01,  0.01)
            (2,0.96) +- (0.01,  0.01)
            (3,0.93) +- (0.01,0.01)
        };
        \legend{
        larger discrepancy,
        smaller discrepancy 
        }

    \end{axis}

\end{tikzpicture}
\caption{Average SSIM according to the instances with larger and smaller delay discrepancies for scenarios with different number of flows in \textit{Random Indoor} Topology.}
\label{fig:ssimClass}
\end{figure}
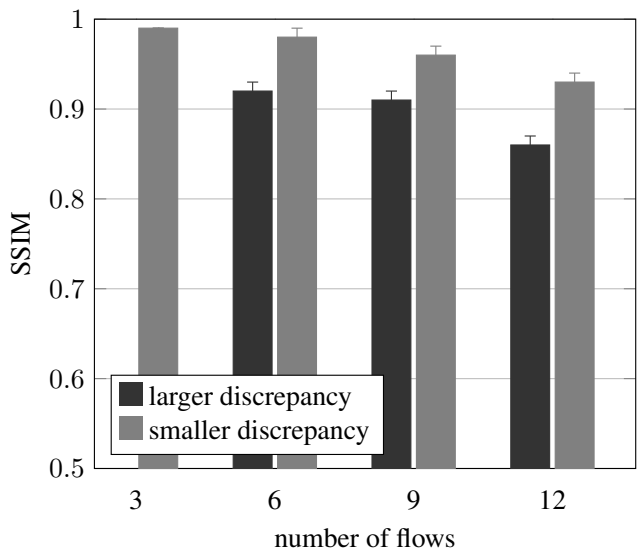

Figure~\ref{fig:ssimClass} plots the average SSIM of the instances for different numbers of flows in the \textit{Random Indoor} topology. According to the delay discrepancy ranges observed in Figure~\ref{fig:gapDelay}, we group experimental run instances in two classes, where the first class exhibits delay discrepancies greater than $100ms$ (called \textit{lager delay discrepancy}) and the second class exhibits delay discrepancies equal to or less than $100ms$ (called \textit{smaller delay discrepancy}) when compared to the results obtained with \textit{ns-3}. From Figure~\ref{fig:ssimClass}, we observe that \textit{larger delay discrepancy} instances were not found in scenarios with 3 flows; however, for the 6-, 9-, and 12-flow experiments, \textit{larger delay discrepancy} instances resulted in lower video quality. Consequently, as previously discussed in Section~\ref{sec_Results}, since MAPE's larger delay discrepancies correlate with poorer video quality,
MAPE estimates could be used to discard paths which would result in inadequate performance for video transmission as they do not currently match the video transmission requirements.
For example, in real-time video applications,
MAPE could be used to identify paths that provide a minimum threshold latency that preserves adequate video quality.

\subsection{Classification Inversions}\label{subsec:RoutingDecisions}
To evaluate how MAPE can be used to inform path selection for video transmission, we use the concept of \textit{classification inversions} ~\cite{passos2018after} defined as follows. Consider two different paths $a$ and $b$. Suppose that video transmissions using path $a$ yield higher SSIM than if path $b$ was used. If MAPE's throughput estimate indicates that path $b$ will outperform path $a$, that constitutes a \textit{classification inversion} using throughput as metric; otherwise, if path $a$ is selected, there is no inversion.

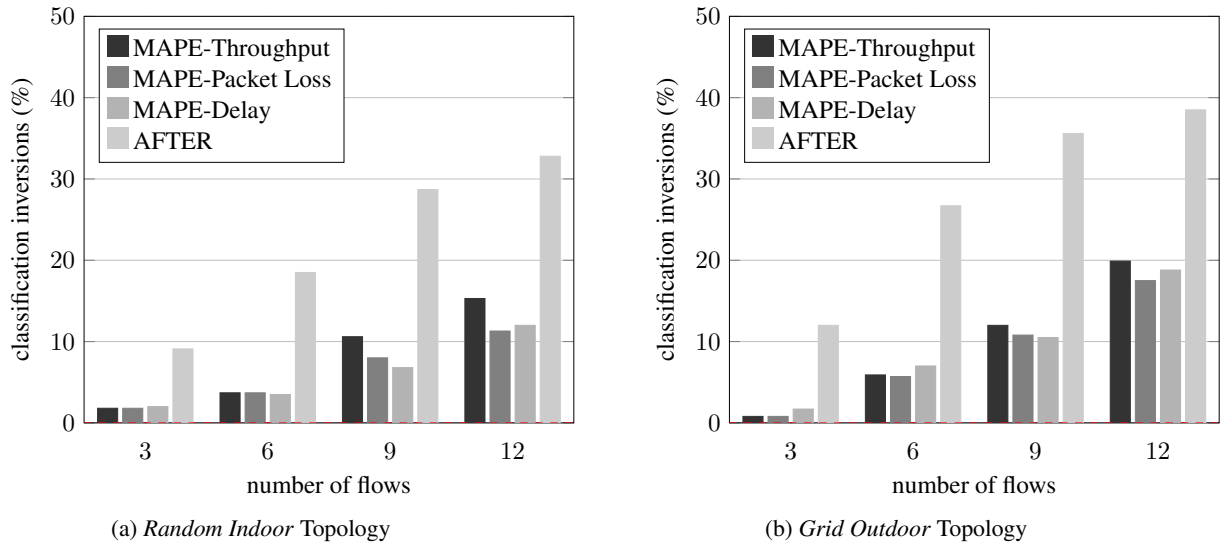
\begin{figure*}[!ht]  
\centering 
  \begin{subfigure}[!ht]{0.4\linewidth}
  \centering 
    \begin{tikzpicture}[scale=0.95]
\pgfplotsset{
    /pgfplots/bar cycle list/.style={/pgfplots/cycle list={
        {black!80,fill=black!80,mark=none},
        {gray,fill=gray,mark=none},
        {gray!60,fill=gray!60,mark=none},
        {gray!40,fill=gray!40,mark=none},
}, },
}
    \begin{axis}[
        ybar,
        bar width=8pt,
        xtick distance=3,
        xlabel=number of flows,
        ylabel=classification inversions (\%),
        enlarge x limits={abs=1.5},
        ymin=0,
        ymax=50,
        xmin=3,
        xmax=12,
        ymajorgrids,
        scaled ticks=false,
        xtick style={
            /pgfplots/major tick length=0pt,
        },
        legend image code/.code={
            \draw[#1, draw=none] (0cm,-0.1cm) rectangle (0.3cm,0.2cm);
        },  
        legend pos=north west,
        legend cell align={left}, 
    ]
        \addplot+ coordinates {
            (3,1.8) +- ( 0,  0)
            (6,3.7) +- (0,  0)
            (9,10.6) +- (0,  0)
            (12,15.3) +- (0,  0)
        };
        \addplot+ coordinates {
            (3,1.8) +- ( 0,  0)
            (6,3.7) +- (0,  0)
            (9,8.0) +- (0,  0)
            (12,11.3) +- (0,  0)
        };
        
        \addplot+ coordinates {
            (3,2.0) +- ( 0,  0)
            (6,3.5) +- (0,  0)
            (9,6.8) +- (0,  0)
            (12,12.0) +- (0,  0)
        };
        \addplot+ coordinates {
            (3,9.1) +- ( 0,  0)
            (6,18.5) +- (0,  0)
            (9,28.7) +- (0,  0)
            (12,32.8) +- (0,  0)
        };
        \addplot+ [draw=red, ultra thin,dashed,smooth] 
    coordinates {(0,0) ( 15,0) };
        
        \legend{
            MAPE-Throughput,
            MAPE-Packet Loss,
            MAPE-Delay,
            AFTER,
        }
    \end{axis}

\end{tikzpicture}
    \caption{\textit{Random Indoor} Topology} \label{fig:inversionSSIM_Random}  
  \end{subfigure}
  \hspace{1.8cm}
\begin{subfigure}[!ht]{0.4\linewidth}
\centering 
   \begin{tikzpicture}[scale=0.95]
\pgfplotsset{
    /pgfplots/bar cycle list/.style={/pgfplots/cycle list={
        {black!80,fill=black!80,mark=none},
        {gray,fill=gray,mark=none},
        {gray!60,fill=gray!60,mark=none},
        {gray!40,fill=gray!40,mark=none},
}, },
}
    \begin{axis}[
        ybar,
        bar width=8pt,
        xtick distance=3,
        xlabel=number of flows,
        ylabel=classification inversions (\%),
        enlarge x limits={abs=1.5},
        ymin=0,
        ymax=50,
        xmin=3,
        xmax=12,
        ymajorgrids,
        scaled ticks=false,
        xtick style={
            /pgfplots/major tick length=0pt,
        },
        legend image code/.code={
            \draw[#1, draw=none] (0cm,-0.1cm) rectangle (0.3cm,0.2cm);
        },  
        legend pos=north west,
        legend cell align={left}, 
    ]
         \addplot+ coordinates {
            (3,0.8) +- ( 0,  0)
            (6,5.9) +- (0,  0)
            (9,12.0) +- (0,  0)
            (12,19.9) +- (0,  0)
        };
        \addplot+ coordinates {
            (3,0.8) +- ( 0,  0)
            (6,5.7) +- (0,  0)
            (9,10.8) +- (0,  0)
            (12,17.5) +- (0,  0)
        };
        
        \addplot+ coordinates {
            (3,1.7) +- ( 0,  0)
            (6,7.0) +- (0,  0)
            (9,10.5) +- (0,  0)
            (12,18.8) +- (0,  0)
        };
        
        \addplot+ coordinates {
            (3,12.0) +- ( 0,  0)
            (6,26.7) +- (0,  0)
            (9,35.6) +- (0,  0)
            (12,38.5) +- (0,  0)
        };
        \addplot+ [draw=red, ultra thin,dashed,smooth] 
    coordinates {(0,0) ( 15,0) };
        
        \legend{
            MAPE-Throughput,
            MAPE-Packet Loss,
            MAPE-Delay,
            AFTER,
        }
    \end{axis}

\end{tikzpicture}
\caption{\textit{Grid Outdoor} Topology} \label{fig:inversionSSIM_Grid}  
\end{subfigure}
\caption{Percentage of classification inversions in terms of SSIM for different numbers of flows.}
\label{fig:inversionSSIM}
\end{figure*} 

To better understand how classification inversions can be used in practice, let us consider the route selection problem in multipath forwarding, where a set of paths needs to be selected for the transmission of multiple flows based on low transmission rate. In this example, the most important aspect is to get the relative ranking of the paths correctly in order to make adequate flow-to-path assignments.

Figure~\ref{fig:inversionSSIM} shows a comparison between MAPE and AFTER in terms of classification inversions for both the \textit{Random Indoor} and the \textit{Grid Outdoor} topologies as a function of the number of flows. 
We evaluate the quality of paths using SSIM metric. 
For MAPE, we consider three possible scenarios: using the average throughput, packet loss or end-to-end delay as metrics to compare the set of paths of all instances. Since AFTER does not estimate delay or packet loss, we only show results when throughput is used to calculate \textit{classification inversions} based on AFTER estimates.
MAPE results in lower percentages of classification inversions (lower than $20\%$) in all scenarios. AFTER, on the other hand, yields $3$ to $6$ times higher classification inversion rates, as it ignores flow transmission rates.

Note that MAPE's packet loss and delay estimates result in lower \textit{classification inversions} for $9$ and $12$ flows when compared to \textit{classification inversions} based on throughput. This demonstrates that both delay and loss should be considered when selecting paths for video transmission, especially when the network becomes saturated. These results are relevant because they confirm that MAPE's estimates, which can be computed in quasi real time, can be used to select paths that improve user QoE in terms of perceived video quality.

\section{Conclusion}\label{sec_Conclusion}
This paper introduced the Multimedia-Aware Performance Estimator, or MAPE for short, a per-flow estimator based on a deterministic discrete event simulation approach. MAPE estimates the throughput, packet loss and end-to-end delay of individual flows using their average transmission rate as input. To the best of our knowledge, MAPE is the first performance estimator that is able to both account for inter-flow interference and accommodate rate-heterogeneous flows, which is essential to more realistically model the behavior of multimedia traffic.

We evaluated MAPE in terms of execution time, prediction accuracy and ability to classify sets of paths according to the video quality at the receiver. Our results indicate that MAPE yields comparable throughput, packet loss and delay estimate accuracy when compared to  stochastic network simulators such as \textit{ns-3} at a fraction of the execution time. When compared to AFTER, through its ability to consider specific per-flow rates, MAPE yields higher accuracy at comparable execution times. We also show in practice that by adopting video coding rates as input, MAPE is able to obtain estimates similar to the ones obtained by ns-3 when driven by  multimedia (MM) traffic. We also demonstrate how MAPE's accurate real-time throughput and delay predictions can be used to make routing decisions for multimedia applications. In particular, we show that MAPE makes correct path selection decisions for over 80\% of the cases, including saturated network scenarios.

As part of future work, we plan to refine MAPE's packet loss and delay models which will help improve its estimation accuracy. While our current implementation uses IEEE 802.11, we also plan to extend MAPE so that it can also be used with other IoT communication technologies such as IEEE 802.15.4 networks. We also intend to further explore the correlation between routing metrics and video quality (e.g., based on the SSIM) and to incorporate into MAPE alternate ways to model variable bitrate streams, including traffic patterns representative of prominent adaptive bitrate streaming traffic, e.g.,  by simulating video frame packets bursts. Our overarching goal is to propose a cross-layer framework that integrates MAPE with video coding for improved QoE.

\bibliographystyle{unsrt}  
\bibliography{references}

\end{document}